\documentclass[aps, prb, reprint]{revtex4-1}

\usepackage[]{graphicx} 
\usepackage{hyperref} 

\draft

\begin{document}

\newcommand{\Vg}[1]{\ensuremath{V_{\mathrm{g#1}}}}
\newcommand{\mug}[1]{\ensuremath{\mu_{\mathrm{g#1}}}}
\newcommand{\Ef}{\ensuremath{E_{\mathrm{F}}}}
\newcommand{\m}[1]{{\bf #1}}

\title{Dopant-controlled single-electron pumping through a metallic island}

\author{Tobias Wenz}
\email{tobias.wenz@ptb.de}
\affiliation{Physikalisch-Technische Bundesanstalt (PTB), Bundesallee 100, 38116 Braunschweig, Germany}
\author{Frank Hohls}
\email{frank.hohls@ptb.de}
\affiliation{Physikalisch-Technische Bundesanstalt (PTB), Bundesallee 100, 38116 Braunschweig, Germany}
\author{Xavier Jehl}
\affiliation{University Grenoble Alpes and CEA-INAC, F-38000 Grenoble, France}
\author{Marc Sanquer}
\affiliation{University Grenoble Alpes and CEA-INAC, F-38000 Grenoble, France}
\author{Sylvain Barraud}
\affiliation{University Grenoble Alpes and CEA-Leti-Minatec, F-38000 Grenoble, France}
\author{Jevgeny Klochan}
\affiliation{Faculty of Physics and Mathematics, University of Latvia, Zellu street 25, LV 1002 Riga, Latvia}
\author{Girts Barinovs}
\affiliation{Faculty of Physics and Mathematics, University of Latvia, Zellu street 25, LV 1002 Riga, Latvia}
\author{Vyacheslavs Kashcheyevs}
\affiliation{Faculty of Physics and Mathematics, University of Latvia, Zellu street 25, LV 1002 Riga, Latvia}

\date{\today}

\begin{abstract}

We investigate a hybrid metallic island\,/\,single dopant electron pump based on fully-depleted silicon on insulator technology. Electron transfer between the central metallic island and the leads is controlled by resonant tunneling through single phosphorus dopants in the barriers. Top gates above the barriers are used to control the resonance conditions. Applying radio frequency signals to the gates, non-adiabatic quantized electron pumping is achieved. A simple deterministic model is presented and confirmed by comparing measurements with simulations.

\end{abstract}

\pacs{}

\maketitle 

Accurate clocked control and transport of single electrons (SE) is a prerequisite for both the emerging field of electron quantum optics\cite{Grenier2011b} and the upcoming introduction of the quantum ampere by fixing the elementary charge~$e$.\cite{Mills2011} Transferring an integer number of electrons~$N$ in each cycle of a drive with frequency $f$ generates a current $I=Nef$. Several approaches to clocked SE transport have been made\cite{Pekola2013} of which tunable-barrier SE pumps are especially promising to reach sufficiently large currents at high accuracy for metrological needs.\cite{Kaestner2015} So far, GaAs based devices achieved the highest verified accuracy at large output currents,\cite{Stein2015,Giblin2012,Bae2015} however a high magnetic field and sub-kelvin temperature are required for accurate operation.

SE pumps made from silicon are promising candidates for less demanding operation requirements and even higher operation frequency\cite{Fujiwara2008} and have been realized using both adiabatic\cite{Jehl2013} and non-adiabatic\cite{Fujiwara2008,Rossi2014} pumping schemes. They benefit from mature CMOS technology with the potential of circuit integration, e.g. on chip drive electronics\cite{Clapera2015} or the charge detectors\cite{Yamahata2011,Tanttu2015} needed for a self-referenced quantized current source.\cite{Fricke2013,Fricke2014} Additionally, CMOS technology offers excellent control of doping, opening the possibility to utilize the large addition energies of single dopants for single-electron control with a potential of increased accuracy. Up to now, the dopant or trap was used as the main quantum dot of a SE pump.\cite{Lansbergen2012,Roche2013,Tettamanzi2014,Yamahata2014} In this letter a new device scheme is examined: Single dopant states located in the barriers are used to control the tunnel coupling of a small metallic island to source and drain. While in other works similar states in barriers were an unwanted side-effect,\cite{Chan2011} here we actively utilize these states to perform SE pumping.

\begin{figure}
  \includegraphics[scale=1]{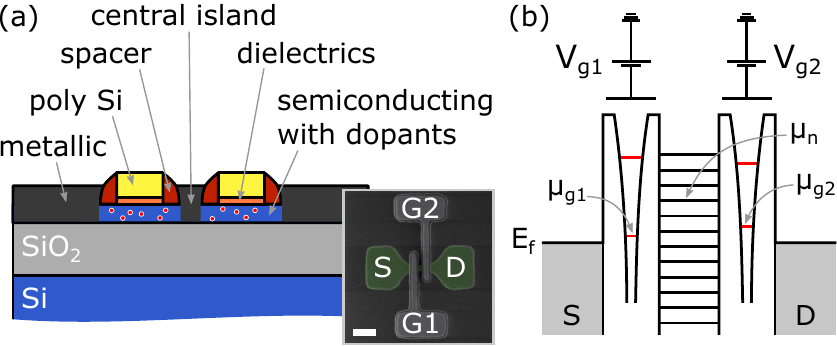}%
  \caption{\label{fig:sketch}(a) Schematic cross section along the nanowire axis. Inset: SEM image of a similar device after gate and spacer patterning (scalebar 100\,nm). (b) Schematic band structure of the investigated device. The central island is metallic. Under each of the gates, single dopants dominate the transport.}%
\end{figure}

The investigated device is produced in a fully CMOS-compatible process using both deep UV and electron beam lithography. It is based on a silicon on insulator wafer with 145\,nm of SiO$_2$, where the bulk silicon layer can be used as a back gate.\cite{Roche2012a} The 20\,nm thick Si-layer is n-doped with phosphorus ($2\cdot10^{18}$\,cm$^{-3}$). A 60\,nm wide nanowire is etched out of the Si-layer. Two top gates (consisting of dielectrics and doped poly Si) are patterned around the nanowire with a length of 60\,nm and a separation of 60\,nm. The gates and a set of 15\,nm wide self-aligned silicon nitride spacers act as a mask for the following doping and silicidation. Consequently, all regions of the nanowire not covered by gate or spacers will be metallic, including a central island formed between the two gates with a width of 30\,nm. A schematic cross section of the device along the nanowire axis is shown in Fig.~\ref{fig:sketch}(a).

The simplified band structure of the device along the length of the nanowire is as follows [Fig.~\ref{fig:sketch}(b)]: Source and drain are metallic. By tuning the top gates to a negative voltage, barriers are formed below the top gates and isolate the metallic region in between the gates from the leads, causing discrete charge states to be formed. Under the gates, the average interdopant distance is approximately $8$\,nm, larger than the Bohr radius $r_{\mathrm{B}}\approx 3$\,nm for P dopants in Si, making these regions semiconducting.\cite{Tabe2010} By choosing the back gate voltage appropriately the device can be tuned such that for each gate only the energy level of a single dopant lies in the transport window and is sufficiently separated from any other state.\cite{Voisin2014, Verduijn2013} The states in the barrier under a single gate can be investigated individually by tuning the gate voltage of the other gate to~$+1$\,V, well above the threshold value. From Coulomb diamond measurements (not shown) the energy separations of the lowest state to the next state are found to be approximately $40$\,meV, comparable to the charging energy commonly found for single phosphorus dopants in Si nanostructures.\cite{Tan2010,Lansbergen2008} States at higher energies are more densely spaced.

\begin{figure}
  \includegraphics[scale=1]{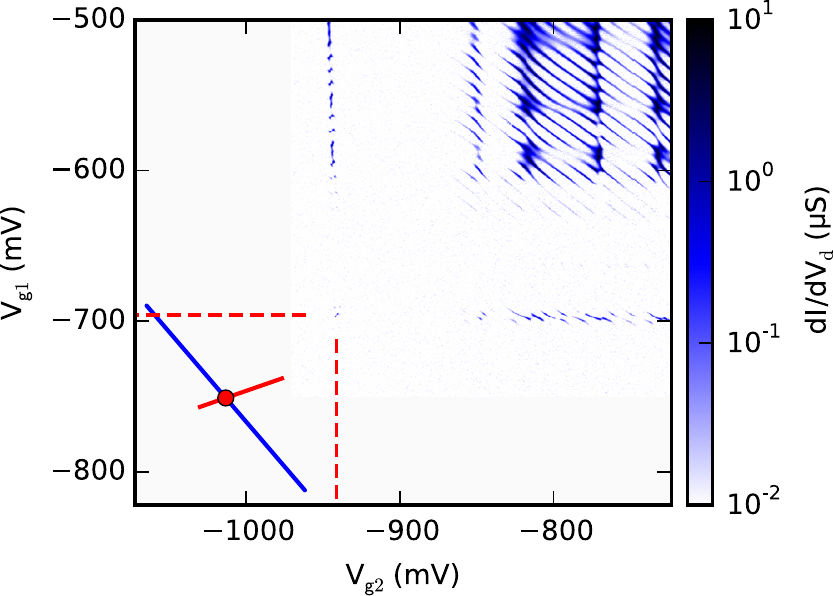}%
  \caption{\label{fig:conductance}Conductance through the device as a function of top gate voltages. Anti-diagonal lines correspond to resonances of the central island. Horizontal and vertical lines are resonances of the states in the barriers. For $\Vg1 < -650$\,mV and $\Vg2 < -890$\,mV, transport through and from the central island is only possible via dopant states (extended with red dashed lines). Plot has been extended beyond measured values to show the line along which the measurement in Fig.~\ref{fig:process}(a) has been taken (red line) and RF amplitude used in that measurement (blue line).}%
\end{figure}

Fig.~\ref{fig:conductance} shows the measured conductance as a function of top gate voltages \Vg1 and \Vg2. All measurements were taken at a base temperature of $T<100$\,mK. The top gates couple to the energy levels of the states in the barriers as well as to the energy levels of the central island. Multiple anti-diagonal lines indicate resonance of the central island with source and drain. They are visible for more positive gate voltages at which the barriers have sufficient transparency. Horizontal (vertical) lines of increased conductance appear, whenever a dopant's state in the barrier under gate~1 (gate~2) is resonant.\cite{Golovach2011} Whenever the anti-diagonal lines intersect a horizontal or vertical resonance line anti-crossings appear, indicating that the central island is capacitively coupled to the states in both barriers. In the absence of these anti-crossings, resonance lines of the central island are evenly spaced as expected for a metallic dot.\cite{Jehl2013} The negligible slope of the horizontal and vertical resonances of the states under the gates demonstrate very small cross-coupling from gate~1 on states located under gate~2 and vice versa, caused by the strong screening provided by the gates and the central island. For more negative gate voltages, conductance is only appreciable in a small region at $\Vg1 = -697$\,mV and $\Vg2 = -947$\,mV, where the first relevant\footnote{The resonance of the lowest state under gate~1 (found at $\Vg1 = -835$\,mV) is very weakly coupled and cannot be resolved in this measurement.} energy level of gate~1 and gate~2 (\mug1 and \mug2, respectively) are both aligned with the Fermi level $\Ef$, i.e. $\mug1=\mug2=\Ef$. These resonances are highlighted with dashed red lines in Fig.~\ref{fig:conductance}.

\begin{figure}
  \includegraphics[scale=1]{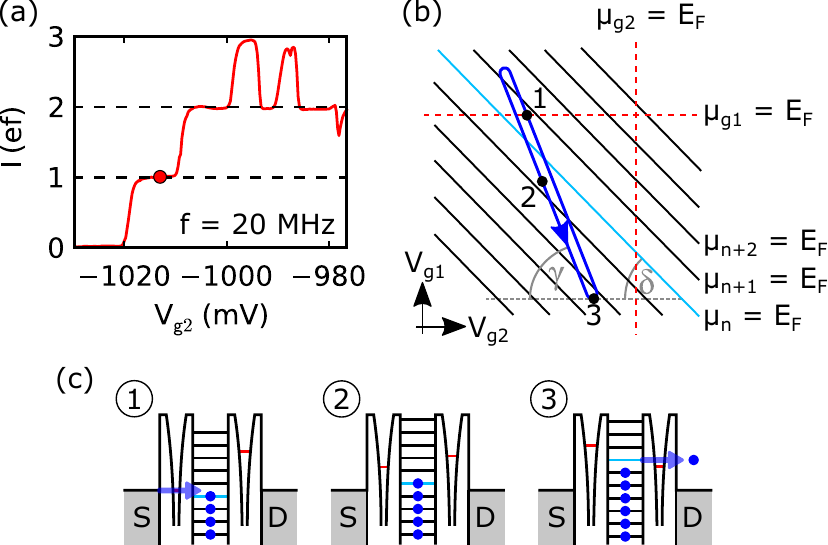}%
  \caption{\label{fig:process}(a) Pumped current as a function of gate voltage along the path indicated in red in Fig.~\ref{fig:conductance}. (b) Stability diagram in the region where transport through the barriers is only possible via the lowest dopant states. Resonance lines of the central island (black) and the dopant states (red) are shown. The path followed by RF voltages is shown in blue (corresponds to blue path indicated in Fig.~\ref{fig:conductance}). This situation corresponds to the case marked with the red dot in (a), where $N=1$ electron is pumped. (c) Positions of energy levels during one cycle of RF modulation.}%
\end{figure}

We will now demonstrate that using the dopant states in the barriers this new type of device allows to generate a quantized current $I=Nef$. Sinusoidal radio frequency (RF) signals with 180 degree phase shift are applied to both gates to follow a certain path in the \Vg1-\Vg2-plane as indicated by the blue line in Fig.~\ref{fig:conductance} ($f=20$\,MHz for all shown measurements). When shifting the center of the blue path (red dot) along the solid red line, the produced current changes, as shown in Fig.~\ref{fig:process}(a). Starting at the most negative voltages, the current increases from $I=0ef$ to $3ef$ in steps of $1ef$, clearly showing quantization of current. Subsequently, more steps between $2ef$ and $3ef$ follow.

To explain the structure of the current quantization steps we consider a simple deterministic model which treats the discrete charge states on the central island and under the barriers as energetically sharp at all times and allows transport only through the resonant states highlighted with red dashed lines in Fig.~\ref{fig:conductance}. The model assumes that the charge of the central island switches if and only if sequential tunneling through a dopant state under the barrier is energetically allowed. Further neglecting the dopant-island capacitive coupling we can reduce the model to the following simple rules. The $n$th electron can only tunnel onto the metallic central island with electro-chemical potential $\mu_n$ if a gate's energy level $\mu_{\mathrm{g}i}$ ($i=1,2$) is in the energy window of the Fermi energy $\Ef$ and the dots potential: $\Ef > \mu_{\mathrm{g}i} > \mu_n$. Similarly, the $n$th electron can only leave the island, if $\Ef < \mu_{\mathrm{g}i} < \mu_n$. The deterministic model is justified for large amplitude and intermediate frequency of modulation, such that on the one hand there is enough time for charge equilibration as the resonance energy passes through the bias window across a dot-lead barrier while on the other hand any off-resonance hopping events remain negligibly rare.

Using these assumptions, we will demonstrate how a quantized current of $N=1$ electron per cycle can be produced; equivalent to the red dot in Fig.~\ref{fig:process}(a). Figure~\ref{fig:process}(b) shows the resonances of energy levels with source and drain as a function of top gate voltages. Red dashed lines represent resonances of the states in the barriers ($\mug1, \mug2 = \Ef$). Black lines represent resonances of the central island's electrochemical potential ($\mu_n=\Ef$), which are at an angle of $\delta\approx38^\circ$ to the \Vg2-axis. For clarity, the $n$ electron resonance is highlighted in cyan. By applying the RF signals to the top gates, the path shown in blue is produced (angle to \Vg2-axis $\gamma=50^\circ$). The RF signals on the gates are $\pi$-shifted, therefore no area is enclosed in the \Vg1-\Vg2-plane. Therefore, this single-parameter pumping process is non-adiabatic.\cite{Kaestner2015}

In Fig.~\ref{fig:process}(c) positions of the energy levels are sketched for certain points along the path. Starting at frame~1, $\mug1\leq\Ef$, allowing the central island to be filled from the source to an occupancy of $n$ electrons. Next, frame~2 highlights the non-adiabaticity of the process:\cite{Kaestner2015} Even though $\Ef<\mu_{n}$, the occupancy of the central island remains unchanged, as no state in any barrier is in the appropriate energy window ($\mu_{n} < \mug1, \mug2$). Finally, in frame~3, $\mug2<\mu_n$, allowing one electron to exit the central island to the drain. Following the path back to frame~1, no electrons enter or leave the dot, as none of the requirements for transport are met. During one cycle, $N=1$ electron has been transported from source to drain in this example.

\begin{figure*}
  \includegraphics[scale=1]{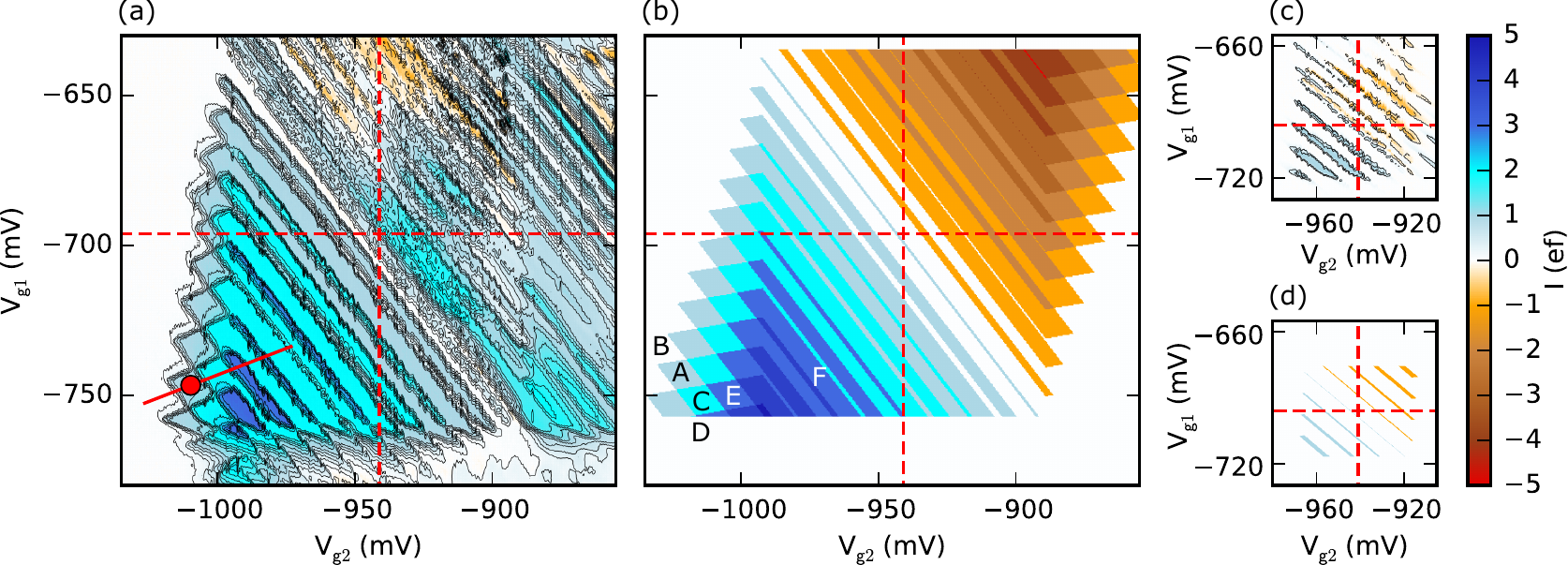}%
  \caption{\label{fig:pumping}(a) Measurement of pumped current as a function of DC gate voltages. Dashed red lines correspond to resonances of states in the barriers under the gates. Solid red line corresponds to the cross section shown in Fig.~\ref{fig:process}(a). (b) Simulation of the produced current using the same parameters as for the measurement in (a) [y-axis shared with (a)]. (c) Measurement of pumped current for smaller amplitude and $\gamma$ closer to $\delta$. Reversal of current can be observed. (d) Simulation corresponding to (c).}%
\end{figure*}

To change the number $N$ of electrons transported per cycle, the top gate DC voltages can be changed. Figure~\ref{fig:pumping}(a) shows a pump current measurement as a function of both top gate voltages, applying the same RF signals as before. The path along which the measurement in Fig.~\ref{fig:process}(a) has been taken is indicated by the solid red line. The situation described in Fig.~\ref{fig:process}(b) is highlighted by the red dot, where a current of $I=1ef$ is measured. Positions in gate voltage of the dopant resonances are marked with red dashed lines. The observed pattern features two main structures: Towards more negative \Vg2, a set of diamonds with quantized current is visible. The first column has a value of $1ef$, the second column a value of $2ef$. Going towards more positive \Vg2, the diamond shape is replaced with a set of anti-diagonal stripes on which the produced current stays constant. Their slope is found to be equal to the angle $\gamma$ of the gate voltage path (blue line in Fig.~\ref{fig:conductance}).

To compare the measurement with the described model and explain the observed changes in the transferred number of electrons per cycle $N$, a simple simulation has been made counting the number of electrons transferred during one pump cycle following the above assumptions for transport to and from the central island. The following experimentally gained parameters are used for the simulation (lever arms $\alpha=\mathrm{d}\mu / e\mathrm{d}V$): The resonance positions of the states in the barriers as function of gate voltages (from Fig.~\ref{fig:conductance}), the lever arms of gate voltage to the barrier state's energy, $\alpha_1\approx 0.26$ and $\alpha_2\approx 0.42$, and to the central island's potential, $\alpha_{1\mathrm{c}}\approx 0.45$ and $\alpha_{2\mathrm{c}}\approx 0.35$, and the charging energy of the central island $E_{\mathrm{c}}\approx 4.3$\,meV (all from Coulomb diamond measurements, not shown). As the tunneling rates are assumed to be infinite, electron transfer takes place whenever possible, and the simulation yields only integer values for $N=I/ef$.

The simulation is shown in Fig.~\ref{fig:pumping}(b). Red dashed lines indicate the resonances of the relevant states under each barrier. The produced pattern is almost point symmetric around the crossing point of those resonances, with the sign of the produced current being flipped for one side of the pattern. First, we will describe the region with positive current which fits well with the measurement. The other half will be discussed later. Point~A is equivalent to the situation depicted in Figs.~\ref{fig:process}(b) and \ref{fig:process}(c), where $N=1$ electron is transmitted. Moving towards more positive \Vg1, point~B is reached, where no electrons are transported. Here, the ejection of the electron in frame~3 can not take place, as the condition for ejection $\mu_n > \mug2$ is not reached. Going from A into the opposite direction towards C the current increases to $N=2$ electrons. Here, an additional electron can tunnel out during frame~3, as $\mu_{n-1} > \mug1$. When \Vg1 is decreased further (point~D), no more electrons are transported. Here, the time dependent gate voltage path does not cross the $\mug1=\Ef$ resonance any more, so no electrons are loaded into the central island during frame~1. When going from point~C to E, an additional electron is loaded onto the central island, as $\mu_{n+2} < \mug1=\Ef$ during frame~1, leading to a total current of $3ef$. Going further towards point~F, the current switches between $4ef$ and $3ef$ several times while the diamond shape is lost. In this region, additional electrons tunnel onto the island during frame~1, but the number of electrons emitted does not increase further, as the $\mug2=\Ef$ resonance is crossed before frame~3. According to the model, the $n$th electron can not be emitted if $\mu_n > \Ef > \mug2$, which gives the anti-diagonal stripes of constant current.

Comparing the simulation in the region where positive current is expected with the measurement, one finds generally good agreement. While the maximum current in the measurement is $3ef$, it is $4ef$ in the simulation. Further, the positive slope of the diamond shapes as well as the slope of the column of diamonds differ. The maximum current and both slopes depend on the lever arms used for the simulation. We have used the values of the latter measured in DC transport with either one of the barriers removed and the potential of the island fixed or both barriers sufficiently transparent even off dopant resonance, which is different from the conditions for pumping. A difference in screening conditions and the neglected capacitive coupling of the central island to the states in the barriers are possible sources of the discrepancy. Capacitive coupling is also likely to cause the zig-zag pattern at the bottom of the measured current map. These effects are subject to further investigation.

When the rules for electron transport are applied to the whole gate range, the simulation suggests that the pattern should be repeated with negative current in the region of more positive gate voltages. This reversal of current is not observed in the data in Fig.~\ref{fig:pumping}(a) due to the presence of additional transport channels that were neglected in the model. These transport channels become more effective at more positive gate voltages. Their effect can be minimized using RF signals with smaller amplitude and $\gamma=40^\circ$ closer to $\delta\approx38^\circ$, yielding a pump current that shows the expected reversal [Fig.~\ref{fig:pumping}(c)], although the produced current is now limited to $I=1ef$. A simulation with a corresponding path shows good agreement [Fig.~\ref{fig:pumping}(d)]. The influence of additional transport channels is subject to further research.

In summary, we have investigated a hybrid quantum dot system consisting of a metallic island, whose tunnel coupling to source and drain is dominated by single dopant states located in the barriers under the top gates. A non-adiabatic pumping scheme using $\pi$-shifted sinusoidal RF signals can be implemented to produce a quantized current which is strongly dependent on the DC gate voltages. With basic assumptions about the system the process can be explained and modeled. A predicted reversal of current, though obscured by additional transport channels at large RF amplitude, is observed for small RF amplitudes. This demonstrates the possibility of gate controlled current reversal for metrological applications. Future work will focus on the reduction of additional resonances in the barriers and on improving the model by including additional transport channels, finite rates and capacitive coupling between dopants and central island in order to investigate the dynamics of coupled single dopant-quantum dot systems.

We are grateful to R. Haug for fruitful discussions. We acknowledge financial support through the EC FP7-ICT initiative under Project SiAM No 610637 and from the Nanosciences foundation through the excellence chair program.

\bibliography{paper}

\end{document}